\documentclass[twoside]{article}
\usepackage{fleqn,espcrc2}
\usepackage[dvips]{graphicx}
\usepackage{epsf}
\usepackage{amssymb}
\def\bc{\begin{center}}
\def\ec{\end{center}}
\def\beq{\begin{equation}}
\def\eeq{\end{equation}}
\def\bs{\begin{slide}}
\def\es{\end{slide}}
\newcommand{\bmath}{\begin{displaymath}}
\newcommand{\emath}{\end{displaymath}}
\newcommand{\beqn}{\begin{eqnarray}}
\newcommand{\eeqn}{\end{eqnarray}}
\newcommand{\beqns}{\begin{eqnarray*}}
\newcommand{\eeqns}{\end{eqnarray*}}
\newcommand{\ba}{\begin{array}{c}} 
\newcommand{\bat}{\begin{array}{cc}} 
\newcommand{\ea}{\end{array}} 
\newcommand{\lef}{(1-\gamma_5)}

\newcommand{\Frac}[2]{\frac{\displaystyle #1}{\displaystyle #2}}

\title{Exclusive hadronic $\tau$ decays in the resonance effective 
action of QCD \thanks{IFIC/00$-$74 report. To appear in the proceedings
of the 6th International
Workshop on Tau Lepton Physics, 18--21 September (2000), Victoria (Canada).}}

\author{J. Portol\'es \address{Departament de F\'{\i}sica Te\`orica, IFIC,
CSIC-Universitat de Val\`encia, \\
Edifici d'Instituts d'Investigaci\'o, Apt. Correus 22085, E-46071 Val\`encia,
Spain}}
       
\begin{document}

\begin{abstract}
Until present the study of the form factors associated to the vector
and axial--vector components of the hadronic weak current has been
carried out with a plethora of modelizations of the underlying strong
interactions. While of importance to get an understanding of the dynamics
involved, the amount and quality of the experimental data start to show
some discrepancies with the analysed models. Moreover, and from a 
theoretical point of view, most of these models are not consistent with
quantum chromodynamics (QCD). We propose a QCD--based model--independent
procedure to analyse those decays through the use of the resonance chiral
theory, the low--energy effective action of QCD in the relevant resonance
region. Within this framework we study the hadronic off--shell width of
the $\rho$ meson and the vector form factor of pion in $\tau^- \rightarrow
\pi^- \pi^0 \nu_{\tau}$. We also comment on the
$\tau^- \rightarrow (\pi \pi \pi)^- \nu_{\tau}$ decay.
\vspace{1pc}
\end{abstract}

\maketitle

\section{Introduction}
The study of matrix elements of hadronic currents in the low--energy 
regime is a long--standing problem of particle physics driven by our
poor knowledge of non--perturbative QCD. In this framework exclusive
hadronic $\tau$ decays ($\tau^- \rightarrow H^- \nu_{\tau}$)
provide an excellent dynamical system to explore,
due to the hadronically clean initial state and the 
factorization between lepton and hadron sectors generically given, in 
the Standard Model, by
\beqn
 M \; & = &  
\Frac{G_F}{\sqrt{2}} \,  V_{CKM} \,  \overline{u}_{\nu} \gamma^{\mu}
\lef  u_{\tau} \,  H_{\mu} \,  ,  \nonumber \\
H_{\mu}  & = &  \langle \,  H \,  | \,  (V_{\mu} - A_{\mu}) 
 e^{i L_{strong}} \, | \, 0 \, \rangle  \, . 
\label{eq:mtau}
\eeqn
Symmetries help us to define a decomposition of $H_{\mu}$ in terms of
the allowed Lorentz structure of implied momenta and a set of 
functions of Lorentz invariants, the {\em form factors} $F_i^H$,
\beq
H_{\mu} \; = \; \sum_i  \underbrace{ \; \,  (  \, \ldots \, )_{\mu}^i 
\; \,}_{Lorentz \, struc.}
 F_i^H (q^2, \ldots) \; .
\label{eq:ff}
\eeq
Form factors are the goal of the hadronic matrix elements evaluation and,
as can be noticed from the definition of $H_{\mu}$ in Eq.~(\ref{eq:mtau}),
are a strong interaction related problem in a non--perturbative regime.
\par
In the last years experiments like ALEPH, CLEO-II, DELPHI and OPAL 
\cite{expe,modl,3pi} have collected an important amount of experimental data
on exclusive channels. Analyses of these data are carried out using
the TAUOLA library \cite{tauo} that includes modelizations of the 
hadronic matrix elements. Though 
heuristically based in expected consequences of QCD, models include 
simplifying assumptions that may be are not well controlled from QCD
itself. Therefore, while of importance to get an understanding of the
dynamics involved, models can be misleading and provide a delusive
interpretation of data. Of particular importance are the processes with 
two and three pseudoscalars in the final state. TAUOLA describes these
using the K\"uhn and Santamaria model \cite{ks}. In order to find out
the size of the hadronic uncertainties hidden in the modelization a 
common procedure is to use another model, typically the Gounaris and
Sakurai \cite{gs}, and then the error is estimated from the difference
between the results of both modelizations \cite{modl}.
\par
The question that we address in this note is how much we can say about 
the semileptonic form factors in a model--independent way. We will study
here the pion vector form factor in $\tau^- \rightarrow \pi^- \pi^0
\nu_{\tau}$ and the problem of defining a hadronic off--shell width
of meson resonances. We comment shortly on the 
$\tau^- \rightarrow (\pi \pi \pi)^- \nu_{\tau}$ decay.

\section{Model--independent knowledge}

Our study is intended to extract information by 
exploiting well--known basic properties of S--matrix theory and QCD.
These provide precise constraints to take into account both on the 
hadronic off--shell widths
of meson resonances and the relevant form factors. We sketch here the
pertinent features.

\subsection{S--matrix theory properties}

On general grounds local causality of the interaction translates into
the analyticity properties of amplitudes and, correspondingly, of
form factors. Being analytic functions in complex variables the 
behaviour of form factors at different energy scales is related and,
moreover, they are completely determined by their singularities. 
Dispersion relations embody rigorously these properties and are the
appropriate tool to enforce them. 
\par
In addition unitarity must be satisfied in all physical regions. This 
S--matrix property provides precise information on the relevant 
contributions to the 
spectral functions of correlators of hadronic currents. These are 
closely related to the form factors.
\par
Unitarity and analyticity complement each other. We will see later on,
in some examples, how these S--matrix theory features help in the 
construction of hadronic observables.

\subsection{QCD}

Though we do not know how to evaluate low--energy hadronic matrix 
elements from QCD itself, a theorem put forward by 
S. Weinberg \cite{wein} and worked out by H. Leutwyler \cite{leu}
sets that, if one writes down the most general
possible Lagrangian, including all terms consistent with assumed
symmetry principles, and then calculates matrix elements with this 
Lagrangian to any given order of perturbation theory, the result
will be the most general possible S--matrix consistent with 
analyticity, perturbative unitarity, cluster decomposition and 
the principles of symmetry that have been specified. It is in this statement
that part of the model--independent work on low--energy hadronic physics
has been based upon.
\par
Massless QCD is symmetric under global independent $SU(N_F)$ rotations
of left-- and right--handed quark fields
\beqn
q_L \, \rightarrow \, h_L \, q_L \, \, & , & \; \; h_L \in SU(N_F)_L , 
\nonumber \\
q_R \, \rightarrow \, h_R \, q_R \, \, & , & \; \; h_R \in SU(N_F)_R ,
\label{eq:rotch}
\eeqn
where $N_F = 2,3$ is the number of light flavours. This is the well--known
$SU(N_F)_L \otimes SU(N_F)_R$ chiral symmetry of QCD 
\cite{wein,chir1,wein1,chir2}.
Quark masses break explicitly this symmetry but what is more relevant is 
that it seems it is also spontaneously broken. Though a rigorous prove
of this feature has only been achieved in the large number of colours
limit \cite{ncinf}, the known phenomenology supports that statement. 
Goldstone theorem demands the appearance of a phase of massless bosons
associated to the broken generators of the symmetry and their quantum
numbers happen to correspond to those of the lightest octet ($N_F =3$) of 
pseudoscalars : $\pi$, $K$ and $\eta$. Their non--vanishing masses are 
due to the explicit breaking of chiral symmetry through quark masses.
\par
This Goldstone phase is fortunate because it provides an energy gap into 
the meson spectrum between the octet of pseudoscalars and the heavier
mesons starting with the $\rho(770)$. We can take precisely the mass of this
resonance $M_{\rho}$ as a reference scale to introduce effective
actions of the underlying QCD: \\ \\
\noindent
a) \underline{\bf{$E \ll M_{\rho}$}}
\vspace*{0.2cm} \\
\hspace*{0.5cm} In this energy region chiral symmetry is the guiding
principle to follow. The relevant effective theory of QCD is chiral
perturbation theory ($\chi$PT) \cite{wein,chir2} that exploits properly
the chiral symmetry $SU(N_F)_L \otimes SU(N_F)_R$. In this effective 
action the active degrees of freedom are those of the octet of 
pseudoscalars and the heavier spectrum has been integrated out. As its
own name implies, $\chi$PT is a perturbation theory in the momenta of
pseudoscalars over a typical scale $\Lambda_{\chi} 
\sim 1 \, \mbox{GeV}$. This entails that the interaction vanishes with the 
momentum, giving an example of dual behaviour between the effective action
(perturbative at low energies) and QCD (where asymptotic freedom prevents
a perturbative expansion in that energy regime). By demanding that 
the interaction satisfies chiral symmetry the complete structure of the 
operators, at a definite perturbative order, is defined. However chiral
symmetry does not give any information on their couplings 
that, in general, carry the information of the contributions of heavier
states that have been integrated out. Their study is outside the scope
of $\chi$PT but some thorough developments have been achieved 
\cite{vmd,nc}.
\par
$\chi$PT is a long--standing successful framework for the study of 
strong and weak processes at very low energies \cite{toni}. \\ \\
\noindent
b) \underline{\bf{$E \simeq M_{\rho}$}}
\vspace*{0.2cm} \\
\hspace*{0.5cm} At $E \simeq 1 \, \mbox{GeV}$ other meson states
(mainly unstables) are active degrees of freedom to take into account.
Chiral symmetry still provides the guide in the construction of the
effective action of QCD, following the pioneering work of S. Weinberg
\cite{wein1} in which the new states are represented by fields that 
transform non--linearly under the axial part of the chiral group.
For the lightest octet of resonances (vectors, axial--vectors,
scalars and pseudoscalars) this procedure was carried out in Ref. 
\cite{vmd} and the resulting effective action is called resonance
chiral theory. As in $\chi$PT, chiral symmetry constraints the
structure of the operators but gives no information on their couplings,
that remain unknown, though they could be studied by using models.
Resonance chiral theory provides, therefore, 
a model--independent parameterization
of the processes involving resonances and pseudoscalars in terms of 
those couplings.  \\ \\
\noindent
c) \underline{\bf{$E \gg M_{\rho}$}}
\vspace*{0.2cm} \\
\hspace*{0.5cm} At much higher energies the asymptotic freedom of QCD
implies that a perturbative treatment of the theory is indeed appropriate. 
The study of this energy region is also important for low--energy
hadron physics because we are able to evaluate, within QCD, the
asymptotic behaviour of spectral functions and then to impose these
constraints on the low--energy regime. This procedure is well supported at
the phenomenological level. \\

To take advantage of the underlying QCD dynamics in the model--independent
study of hadronic observables, we have to consider the entire 
view and how different energy regimes intertwine between themselves.

\section{Hadronic off--shell width of meson resonances}

Before addressing the issue of form factors in exclusive hadronic 
$\tau$ decays let us consider one of its basic elements : the off--shell 
width of meson resonances. We will focus in the width
of the $\rho$ though the discussion is analogous for any meson 
resonance. In Ref. \cite{pg97} it was pointed out
that one could consider the evaluation of the off--shell $\rho$ width
from the resonance chiral theory.
The detailed account, that I shortly report here, has been developed
in Ref. \cite{gpp00}.
\par
The resonance chiral effective theory with three flavours and 
including only vector meson resonances is given, at the lowest chiral
order, by \cite{vmd}
\beq
{\cal L}_{\chi V} \, = \, {\cal L}_{\chi}^{(2)} \, + \,
{\cal L}_{KV} \, + \, {\cal L}_V^{(2)} \; .
\label{eq:lcr}
\eeq
Here ${\cal L}_{\chi}^{(2)}$ is the ${\cal O}(p^2)$ chiral Lagrangian
\beq
{\cal L}_{\chi}^{(2)} \, = \, \Frac{F^2}{4} \, \langle \, u_{\mu}
u^{\mu} \, + \, \chi_+ \, \rangle \; ,
\label{eq:lc}
\eeq
where $F$ is the pion decay constant ($F \approx 92.4 \, \mbox{MeV}$),
\begin{eqnarray}
u_{\mu} \, & = & \, i \, [ u^{\dagger} (\partial_{\mu} - i r_{\mu}) u
\, - \, u ( \partial_{\mu} - i \ell_{\mu}) u^{\dagger} \, ] \; , 
\nonumber \\
 \chi_{+} \, & = &  \, u^{\dagger} \chi u^{\dagger} \, + \,
u \chi^{\dagger} u \; ,  \nonumber \\
u \,  & = &   \, \exp \left( \Frac{i}{\sqrt{2}F} \Pi \right) \, ,
\label{eq:forma}
\end{eqnarray}
and $\Pi$ is the usual representation of the Goldstone fields
\beq
\Pi  =  \left( \begin{array}{ccc}
                   \Frac{\pi^0}{\sqrt{2}} \, + \,
           \Frac{\eta_8}{\sqrt{6}} & \pi^+ & K^+ \\
           \pi^- & - \Frac{\pi^0}{\sqrt{2}} \, + \,
           \Frac{\eta_8}{\sqrt{6}} & K^0 \\
           K^- & \overline{K^0} & - \Frac{2}{\sqrt{6}} \eta_8
           \end{array}
            \right) .
\label{eq:pi}
\eeq
In Eq.~(\ref{eq:lc}), $\langle A \rangle$ stands for a trace in
flavour space.
${\cal L}_{\chi}^{(2)}$ has a $SU(3)_L \otimes SU(3)_R$ chiral gauge
symmetry supported by the external fields $\ell_{\mu}$, $r_{\mu}$
and $\chi$.
\par
In Eq.~(\ref{eq:lcr}), ${\cal L}_{KV}$ is the kinetic Lagrangian of vector
mesons and ${\cal L}_V^{(2)}$ describes the chiral couplings of
vector mesons to the Goldstone fields and external currents at the
lowest order,
\beq
{\cal L}_V^{(2)} \, = \, \Frac{F_V}{2 \sqrt{2}} \, \langle V_{\mu \nu}
\, f_+^{\mu \nu}  \rangle \, + \, i  \Frac{G_V}{\sqrt{2}} \,
\langle  V_{\mu \nu} \, u^{\mu} \, u^{\nu}  \rangle \, ,
\label{eq:l2v}
\eeq
where $f_+^{\mu \nu} \, = \, u F_L^{\mu \nu} u^{\dagger} \, + \,
u^{\dagger} F_R^{\mu \nu} u$, with $F_{L,R}^{\mu \nu}$ the field
strength tensors of the left and right external currents $\ell_{\mu}$ and
$r_{\mu}$. $V_{\mu \nu}$
denotes the octet of the lightest vector mesons, in the antisymmetric
formulation \cite{chir2,vmd}, with a flavour content analogous to $\Pi$ in
Eq.~(\ref{eq:pi}). The effective couplings $F_V$ and $G_V$ can be
determined from the decays $\rho^0 \rightarrow e^+ e^-$ and
$\rho^0 \rightarrow \pi^+ \pi^-$ respectively.
Notice that only linear terms in the vector fields have been considered in
${\cal L}_V^{(2)}$.
\par
Assuming unsubtracted dispersion relations for the pion and axial
form factors one gets two constraints \cite{vmd1} among the couplings
$F$, $F_V$ and $G_V$:
\begin{eqnarray}
F_V \, G_V \; = \; F^2 \; , \; \; & \; \; & \; \; \;
F_V \, = \, 2 \, G_V \; ,
\label{eq:const}
\end{eqnarray}
which are reasonably well satisfied phenomenologically.
We enforce those constraints in our analysis.
\par
A Lagrangian density that includes the Goldstone
bosons and spin--1 resonances is not unique
but depends on the definitions of the fields, whereas the observable
physical quantities should be
independent of them. To take into account this feature and in order 
to construct the dressed
propagator of the $\rho$ meson we should consider, for a definite 
intermediate state, all the
contributions carrying the appropriate quantum numbers. The first cut,
in the $\rho$ case, is a two--pseudoscalar absorptive 
contribution that happens to saturate its width. Here we will
take into account two--particle absorptive contributions only.
In order to fulfil short--distance QCD constraints, effective
vertices are not only diagrammatically driven by the $\rho$ 
propagator but also through local contributions.
\begin{figure}[htb]
\vspace*{-0.7cm}
\includegraphics*[scale=0.65]{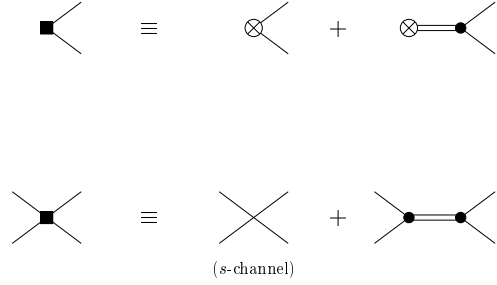}
\vspace*{-1.4cm}
\caption{Effective vertices contributing to vector transitions in the
s--channel. The crossed circle stands for an external vector current
insertion. A double line indicates the $\rho$ and single ones the 
pseudoscalars. Local vertices on the right--hand side are provided,
at leading order, by ${\cal L}^{(2)}_{\chi}$.}
\end{figure}
\par
The construction of these vertices goes as sketched in Figure 1
where, at the leading order, the local vertices on the right--hand side
of the equivalence are provided by the ${\cal O}(p^2)$ chiral
Lagrangian ${\cal L}_{\chi}^{(2)}$ in Eq.~(\ref{eq:lc}). The diagrams
contributing to physical observables will be constructed taking into
account all possible combinations of these two effective vertices.
\par
We propose to define the spin--1 meson width
as the imaginary part
of the pole generated by resumming those diagrams, with an absorptive part
in the s--channel, that contribute to the two--point function of the
corresponding vector current. That is, the pole
of
\beq
\Pi_{\mu \nu}^{jk} \, = \, i \, \int \, d^4 x \, e^{iqx} \,
\langle 0 | \, T (V_{\mu}^j(x) V_{\nu}^k(0)) \, | 0 \rangle \; ,
\label{eq:two}
\eeq
with
\beq
V_{\mu}^j \, = \, \Frac{\delta S_{\chi V}}{\delta v_j^{\mu}} \; ,
\label{eq:curr}
\eeq
with $S_{\chi V}$ the action associated to ${\cal L}_{\chi V}$.
We take the $\rho^0$ quantum numbers that correspond to $j=k=3$.
Lorentz covariance and current conservation
allow us to define an invariant
function of $q^2$ through
\begin{eqnarray}
\Pi_{\mu \nu}^{33} \, & = &  \, (q^2 g_{\mu \nu} \, - \, q_{\mu} q_{\nu}) \,
\Pi^{\rho}(q^2) \; , \nonumber \\ \, \label{eq:inva}  \\
\Pi^{\rho}(q^2) \, & = & \, \Pi_{(0)}^{\rho} \, + \, \Pi_{(1)}^{\rho} \, + \,
\Pi_{(2)}^{\rho} \, + \cdots  \; , \nonumber
\end{eqnarray}
where $\Pi_{(0)}^{\rho}$ corresponds to the tree level contribution,
$\Pi_{(1)}^{\rho}$ to the one--loop amplitudes and so forth.
By evaluating those diagrams in Figure 2 we find
\beqn
\Pi^{\rho}(q^2) \, & = & \,  \Frac{F_V^2}{M_V^2 - q^2} \, \left[
 1  +  \omega  \sum_{n=0}^{\infty} \, \left( \, \Frac{q^2}{M_V^2} \,
\omega  \right)^n \right] \; , \nonumber \\
\omega \, & = &\, - \, \Frac{M_V^2}{F_V^2} \, \Frac{M_V^2}{M_V^2 - q^2} \,
4 \, \overline{B_{22}} \; ,
\label{eq:omiga}
\eeqn
where $\overline{B_{22}}$ has been defined properly in Ref. \cite{gpp00}.
\begin{figure}[htb]
\vspace*{-0.8cm} 
\includegraphics*[scale=0.7]{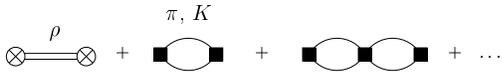}
\vspace*{-1cm}
\caption{Dyson--Schwinger resummation giving $\Pi^{\rho}(q^2)$. The
effective squared vertices are those given in Figure 1.}
\vspace*{-0.1cm}
\end{figure}
\par
Now, resumming, using that $F_V^2 = 2 F^2$ (see Eq.~(\ref{eq:const})),
we get for the imaginary part of the pole
\beqn
M_V \Gamma_{\rho}(q^2) \, & = & \, - \, 2  M_V^2 \, \Frac{q^2}{F^2} 
\, \mbox{Im} \, \overline{B_{22}} \nonumber \\
\, & = &
\Frac{M_V^2 \, q^2}{96 \, \pi \, F^2} \,
\biggl[ \,  \sigma_{\pi}^3 \, 
\theta(q^2 - 4 m_{\pi}^2) \, \biggr. \nonumber \\
 & \, & \; \; \; \; \; \left. +  \Frac{1}{2}  \sigma_K^3 \,
\theta(q^2 - 4 m_{K}^2)   \right] \, ,
\label{eq:rhooff}
\eeqn
where $ \sigma_P \, = \, \sqrt{1-4 m_P^2/q^2} $.
In Ref. \cite{gpp00} it is shown that this result is independent of the
spin--1 field formulation that has been chosen. We emphasize that our 
result for $\Gamma_{\rho}(q^2)$ is constrained by chiral symmetry
and the asymptotic behaviour of $\mbox{Im} \, \Pi^{\rho}(q^2)$ ruled
by QCD. It would be demanding to have a different, leading, off--shell
behaviour without spoiling those all--important features.
\par
A similar procedure could be
applied to higher multiplicity intermediate states by constructing
the relevant effective vertices analogous to those in Figure 1. Although
technically much more involved, its study would be necessary to evaluate
the width of other resonances such as the $\omega$(872) or the 
axial--vector mesons.

\section{Vector form factor of the pion}

The pion vector form factor, $F_V(s)$ is defined through
\beq
\langle \, \pi^+(p') \, \pi^-(p) \, | \, V_{\mu}^3 \, | \, 0 \, \rangle
\, = \, (p-p')_{\mu} \, F_V(s) \; , 
\label{eq:pff}
\eeq
where $s=q^2=(p+p')^2$ and $V_{\mu}^3$ is the third component of the 
vector current associated to the $SU(3)$ flavour symmetry of the QCD
Lagrangian. This form factor drives the hadronic part of both 
$e^+ e^- \rightarrow \pi^+ \pi^-$ and $\tau^- \rightarrow \pi^- \pi^0 \nu_{\tau}
$ processes in the isospin limit. At very low energies, $F_V(s)$ has been
studied in the $\chi$PT framework up to ${\cal O}(p^6)$ \cite{fv1,fv2}. A
successful study at the $\rho(770)$ energy scale has been carried out
in the resonance chiral theory in Ref. \cite{pg97}.
\par
Analyticity and unitarity properties of $F_V(s)$ tightly constrain, on general
grounds, the structure of the form factor \cite{pg97,noi}. Elastic unitarity
and Watson final--state theorem relate the imaginary part of $F_V(s)$ to 
the partial wave amplitude $t_1^1$ for $\pi \pi$ elastic scattering, with
angular momentum and isospin equal to one, as
\beq
\mbox{Im} \, F_V(s+i\varepsilon) \, = \; e^{i \delta_1^1} \, \sin (\delta_1^1) 
\, F_V(s)^* \, , 
\eeq
that shows that the phase of $F_V(s)$ must be $\delta_1^1$. Thus analyticity
and unitarity properties of the form factor are accomplished by demanding
that it should satisfy a n--subtracted dispersion relation with the 
Omn\`es solution \cite{pg97,omn}
\beqn
F_V(s)  & = &  \exp \left\{ \sum_{k=0}^{n-1} \Frac{s^k}{k!} 
\Frac{d^k}{ds^k} \ln F_V(s)|_{s=0} \; \right. \nonumber \\
& & \; \; \; \; \; \; \; \left. + \, \Frac{s^n}{\pi}
 \int_{4 m_{\pi}^2}^{\infty} 
\Frac{dz}{z^n} \Frac{\delta_1^1 (z)}{z - s - i \varepsilon} \right\} .
\label{eq:solom}
\eeqn
This solution is strictly valid only below the inelastic threshold 
($s  <  16 m_{\pi}^2$), however higher multiplicity intermediate states
are suppressed by phase space and ordinary chiral counting. The
$\delta_1^1(s)$ phase--shift, appearing in Eq.~(\ref{eq:solom}), is
rather well known, experimentally, up to $E \sim 2 \, \mbox{GeV}$
\cite{ochs}. 
\par
Therefore and with an appropriate number of subtractions we can parameterize
$F_V(s)$ with the subtraction constants appearing in the first term of the
exponential in Eq.~(\ref{eq:solom}). In Ref. \cite{noi} we have used
three subtractions :
\beqn
F_V(s)  & = &  \exp \left\{ \alpha_1 s \, + \, \Frac{1}{2} \alpha_2
s^2 \, \right. \nonumber \\
& & \; \; \; \; \; \; + \left. \Frac{s^3}{\pi} \int_{4 m_{\pi}^2}^{\Lambda^2}
\Frac{dz}{z^3} \Frac{\delta_1^1(z)}{z - s - i \varepsilon} \, \right\} ,
\label{eq:sol3}
\eeqn
where we have introduced an upper cut in the integration, $\Lambda$. 
This cut--off has to be taken high enough not to spoil the, a priori, 
infinite interval of integration, but low enough that the integrand  is
well known in the interval. The two subtraction constants $\alpha_1$ and
$\alpha_2$ (a third one is fixed by the normalization $F_V(0) = 1$) are
related with the squared charge radius of the pion 
$\langle r^2 \rangle_V^{\pi}$ and the quadratic term $c_V^{\pi}$ in the
low--energy expansion 
\beq
F_V(s) \, = \, 1 + \Frac{1}{6} \langle r^2 \rangle_V^{\pi} s \, + \,
c_V^{\pi} s^2 \, + \, {\cal O}(s^3) ,
\label{eq:chirex}
\eeq
through the relations
\beqn
\langle r^2 \rangle_V^{\pi} &  = & 6 \, \alpha_1 \; , \nonumber \\
c_V^{\pi} & = & \Frac{1}{2} ( \alpha_2 + \alpha_1^2 ) \, .
\label{eq:allow}
\eeqn
$F_V(s)$ endows the hadronic dynamics in the 
$\tau^- \rightarrow \pi^- \pi^0 \nu_{\tau}$ decay. This process has
recently been measured accurately by three experimental groups \cite{modl} :
ALEPH, CLEO-II and OPAL. We have taken $F_V(s)$,
as given by Eq.~(\ref{eq:sol3}), to fit the ALEPH set of data. The input
of the $\delta_1^1(s)$ phase--shift is included as follows \cite{noi}. 
Resonance chiral theory and vector meson dominance provide a 
model--independent analytic expression that describes properly the
$\rho(770)$ contribution \cite{pg97}
\beq
\delta_1^1(s) \, = \, \arctan \left\{ \Frac{M_{\rho}
\Gamma_{\rho}(s)}{M_{\rho}^2 - s} \right\} \, ,
\label{eq:d11r}
\eeq
with $\Gamma_{\rho}(s)$ the off--shell $\rho(770)$ width in 
Eq.~(\ref{eq:rhooff}). This phase--shift is accurate up to 
$E \sim 1 \, \mbox{GeV}$. At higher energies heavier resonances with the
same quantum numbers pop up and we use the available experimental data
from Ochs \cite{ochs}. However there are still contributions that are not
taken into account with Ochs data. These are those of coupled channels
that open at the $K \overline{K}$ threshold \cite{pal}. Therefore
in order to have
a conservative determination of the observables we choose to fit the
ALEPH data up to $\sqrt{s} \simeq 1.1 \, \mbox{GeV}$ where we have a 
thorough control of the contributions. We obtain, using 
Eq.~(\ref{eq:allow}),
\beqn
M_{\rho} & = &  (0.775 \, \pm \, 0.001) \, \mbox{GeV}, \nonumber \\
\langle r^2 \rangle_{V}^{\pi} & = & (11.04 \, \pm \, 0.12) \,  \mbox{GeV}^{-2} , 
\label{eq:res1}  \\
c_V^{\pi} & = &  (3.78 \, \pm \, 0.04) \,  \mbox{GeV}^{-4} , \nonumber
\eeqn
with $\chi^2/d.o.f. = 33.8/21$. 
In Figures 3 and 4 we show the prescription of $F_V(s)$ as given by
Eq.~(\ref{eq:sol3}) with the results of the fit
(Eq.~(\ref{eq:res1})).

\begin{figure}[htb]
\vspace*{-0.9cm}
\hspace*{-0.7cm} 
\includegraphics*[angle=-90,scale=0.55]{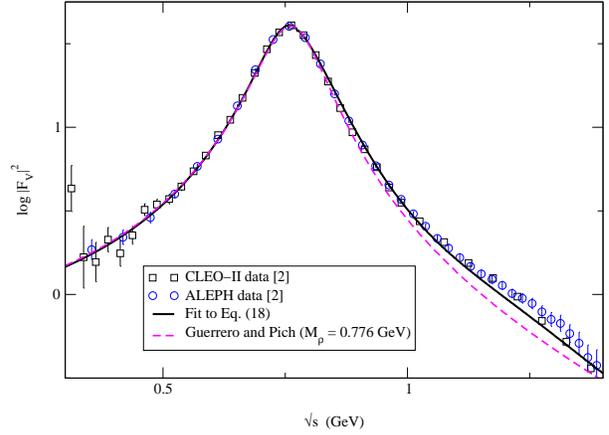}
\vspace*{-1.45cm} 
\caption{Comparison of experimental data of $\tau$ decays with the
results of our fit (Eq.~(\protect\ref{eq:sol3})) and the prediction
of the Guerrero and Pich form factor (Eq.~(\protect\ref{eq:fvfino})).}
\vspace*{-0.4cm}
\end{figure}
\begin{figure}[htb]
\vspace*{-1cm}
\hspace*{-0.8cm} 
\includegraphics*[angle=-90,scale=0.505]{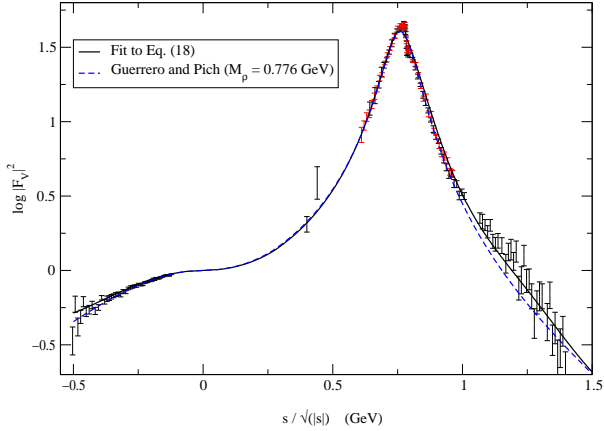}
\vspace*{-1.45cm} 
\caption{Comparison of experimental data of $e^+ e^-$ (time--like)
\protect\cite{ee} and $e \pi$ (space--like) \protect\cite{epi} with the
results of our fit to $\tau$ data (Eq.~(\protect\ref{eq:sol3})) and 
the prediction
of the Guerrero and Pich form factor (Eq.~(\protect\ref{eq:fvfino})).}
\vspace*{-0.7cm}
\end{figure}
A different model--independent procedure was followed
in Ref. \cite{pg97} where $F_V(s)$ was compelled by further requirements
than those of analyticity and unitarity. The behaviour of the
vector form factor of the pion, at low energies, is driven by chiral 
symmetry. The $SU(2)$ result of $\chi$PT up to ${\cal O}(p^4)$ \cite{fv1} 
is 
\beq
F_V^{\chi}(s)  =  1  +  \Frac{2 L_9^r(\mu)}{F^2} s 
 -  \Frac{s}{96 \pi^2 F^2}  A \left( \Frac{m_{\pi}^2}{s} ,
\Frac{m_{\pi}^2}{\mu^2} \right) , 
\label{eq:1loop}
\eeq
where $L_9$ is an, a priori unknown, coupling from the ${\cal O}(p^4)$
$\chi$PT Lagrangian, and
 $A(x,y) = \ln y + 8 x - 5/3 +  
\sigma_x^3 \ln ((\sigma_x+1)/(\sigma_x-1))$, $\sigma_x = \sqrt{1-4x}$, 
the two--point one--loop integral containing the chiral logarithms.
Alternatively one could consider the evaluation of $F_V(s)$ in the 
leading large $N_C$ limit where the vector form factor is given by
an infinite set of resonant contributions. Considering only the
$\rho(770)$ contribution we have \cite{vmd1}
\beq
F_V^{N_C}(s) = \Frac{M_{\rho}^2}{M_{\rho}^2 - s} \, .
\label{eq:ncinf}
\eeq
Guerrero and Pich proceeded by matching both results $F_V^{\chi}(s)$ and
$F_V^{N_C}(s)$ . This procedure is simplified because of the fact that 
the one--loop contribution to $F_V^{\chi}(s)$ is next-to-leading
in the $1/N_C$ expansion. Then we have
\beq
F_V(s) = \Frac{M_{\rho}^2}{M_{\rho}^2-s} - \Frac{s}{96 \pi^2 F^2}
A \left( \Frac{m_{\pi}^2}{s}, \Frac{m_{\pi}^2}{M_{\rho}^2} \right).
\label{eq:match}
\eeq
Notice that this result implies a resummation of the polynomial part 
given by $\chi$PT and prescribes a value for $L_9^r(M_{\rho})$. The
next step was to match this result with the prescription provided
by analyticity and unitarity through the Omn\`es solution 
(\ref{eq:solom}). This gives 
\beq
F_V(s) \! = \! \Frac{M_{\rho}^2}{M_{\rho}^2-s} \exp \left\{ 
\Frac{-s}{96 \pi^2 F^2} A \left( \Frac{m_{\pi}^2}{s} , 
\Frac{m_{\pi}^2}{M_{\rho}^2} \right) \! \right\} . 
\label{eq:expo}
\eeq
Finally, noticing that the pion contribution to $\Gamma_{\rho}(s)$
in Eq.~(\ref{eq:rhooff}) can be written too as
\beq
\Gamma_{\rho}(s) = - \Frac{M_{\rho} s}{96 \pi^2 F^2} \, 
\mbox{Im} \, A \left( \Frac{m_{\pi}^2}{s} , 
\Frac{m_{\pi}^2}{M_{\rho}^2} \right) , 
\label{eq:wir}
\eeq
Guerrero and Pich moved this piece from the exponential to the pole, with
the final result
\beqn
F_V(s) \! \! & \! = \! \! \! & \! \Frac{M_{\rho}^2}{M_{\rho}^2 -s - i M_{\rho} 
\Gamma_{\rho}(s)}  \times \nonumber \\
& & \! \!  \exp \left\{ \Frac{-s}{96 \pi^2 F^2} \mbox{Re} 
A \left( \Frac{m_{\pi}^2}{s} , 
\Frac{m_{\pi}^2}{M_{\rho}^2} \right) \! \right\}.
\label{eq:fvfino}
\eeqn
Note that the only free parameter in this expression is $M_{\rho}$. 
In spite of including the $\rho(770)$ resonance only, and as can be
seen in Figures 3 and 4, it does an excellent
job in describing experimental data up to $E \sim 1 \, \mbox{GeV}$,
showing, once again, the compelling role of chiral symmetry, vector 
meson dominance, analyticity and unitarity in the description of form
factors.

\section{$\tau^- \rightarrow (3 \pi)^- \nu_{\tau}$ : axial--vector form
factors}

Similar ideas to the ones proposed previously can be applied to more
complicated processes. In $\tau^- \rightarrow (\pi \pi \pi)^- \nu_{\tau}$
the hadronic matrix element has, in the isospin limit, contribution 
from the corresponding axial--vector current ($A_{\mu}^-$) only. These
form factors can be defined as
$$
\! \! \! \! \! \! \! \! \! \! \! \! \! \! \! \! \! \! \! \! \! \! \! \! \!
 \! \! \! \! \! \! \! \! \! \! \! \! \! \!
\langle \, \pi^-(p_1) \, \pi^-(p_2) \, \pi^+(p_3) \, |  A_{\mu}^-  |\,  0 \,
 \rangle  \, = 
$$
\vspace*{-0.5cm}
\beqns
 \left( g_{\mu \nu} - \Frac{Q_{\mu} Q_{\nu}}{Q^2} \right) 
\,  \left[ \, F_1(Q^2,s_1,s_2) \, (p_1-p_3)^{\nu} \right. 
 & &  \nonumber \\
  + \,  \left. F_1(Q^2,s_2,s_1) \, (p_2-p_3)^{\nu}
  \, \right] \mbox{$ \! \! \! \! \! \! \! \!$} & & 
\eeqns
\beq
 \; \; \; \; + \, F_2(Q^2,s_1,s_2) \, Q_{\mu}   \, ,
\label{eq:axialf}
\eeq
where  $Q= p_1+p_2+p_3$, $s_i = (Q-p_i)^2$. In the chiral limit ($m_{\pi}=0$)
$F_2(Q^2,s_1,s_2) = 0$ and we have just one form factor $F_1(Q^2,s_1,s_2)$
to work out. This form factor is driven by both vector and axial--vector
resonances.
\par
A thorough analysis of $F_1(Q^2,s_1,s_2)$ in the model--independent framework
of resonance chiral theory and vector meson dominance is under way 
\cite{noi2}. Let us just to point out here the necessity of this study. As 
we said in the Introduction the present sets of experimental data are analysed
using the TAUOLA library where the K\"uhn and Santamaria \cite{ks} hadronic
matrix elements are implemented.  This model was constructed to be
consistent with $\chi$PT
at the leading ${\cal O}(p^2)$ order and
with the asymptotic behaviour that QCD demands to form factors. 
At ${\cal O}(p^4)$ in $\chi$PT the contribution of spin--1 resonances 
appears, and the 
low--energy behaviour of $F_1(Q^2,s_1,s_2)$ given by the model is
\beq
F_1^{KS} \,   \simeq \,  1 \,   + \,   
\Frac{s_1}{M_V^2}  \, + \,  \Frac{Q^2}{M_A^2} \, +  \, {\cal O} \left(
\Frac{q^4}{M_R^4} \right) \, ,
\label{eq:ksexp}
\eeq
where $M_V$ and $M_A$ are masses of vector and axial--vector resonances,
respectively. However the evaluation of the form factor within resonance
chiral theory \cite{noi2} provides the low--energy expansion result
\beq
F_1^{\chi} \,   \simeq \,  1 \,   + \,   \Frac{3}{2} \, 
\Frac{s_1}{M_V^2}  \, + \,  (\ldots)  \, +  \, {\cal O} \left(
\Frac{q^4}{M_R^4} \right) \, ,
\label{eq:rctexp}
\eeq
in agreement with the evaluation within the framework of $\chi$PT at
${\cal O}(p^4)$ \cite{urech}. This shows that the K\"uhn and Santamaria
model is no longer consistent, at ${\cal O}(p^4)$, 
with the chiral symmetry of QCD and, therefore, further studies are required.
\vspace*{-0.15cm} \\
{\bf Acknowledgements}  \\
I wish to thank D. G\'omez Dumm and A. Pich for a fruitful and demanding
collaboration, and to R.J. Sobie and J.M. Roney for the luxurious organization
of the Tau2000 meeting. I thank also A. Pich for a critical reading of 
the text. This work has been partially supported by CICYT
PB97--1261.

\end{document}